# Exact Stark analytical function for H$_\alpha$ line based on the FFM Model related with plasma parameters

A. Sarsa[1], A. Jiménez-Solano[1], M.S. Dimitrijević[2], C. Yubero[1]*

[1] Departamento de Física, Campus de Rabanales Edif. C2, Universidad de Córdoba, E-14071 Córdoba, Spain

[2] Astronomical Observatory, Volgina 7, 11060 Belgrade, Serbia.

*Corresponding author:* f62yusec@uco.es

**Abstract**

Optical Emission Spectroscopy is a widely used technique for plasma diagnosis, with particular interest in hydrogen atomic emission due to its prevalence in plasmas. However, accurately determining plasma parameters like electron density, electron temperature, and gas temperature starting from the experimental profiles remains a challenge. This paper introduces a comprehensive model for Stark broadening of the Hα line in a wide range of plasma conditions, addressing the limitations of existing analytical expressions for line shapes. The proposed model encompasses the full splitting of the transition into fifteen Lorentzian profiles and electric micro-field fluctuations surrounding the emitting atoms due to collisions with charged particles. Starting from accurate spectral data obtained from realistic computer simulations, fitting parameters of the model, have been obtained by using an optimization method based on a genetic algorithm. The set of parameters of the model are reported for a wide range of plasma conditions. The behavior of these parameters is analyzed to understand their dependence in terms of the electron density and temperature and gas density of the plasma. The model parameters here obtained constitute a useful tool in plasma diagnosis to obtaining the values of the physical parameters of the plasma starting from the experimental profiles.

# 1. Introduction

Plasmas provide an ideal environment for reactions of interest for a variety of applications including gas detoxification, materials processing (thin film deposition, surface functionalization…), plasma catalysis, elemental analysis, treatment of liquids (water) with plasma and sterilization [1-6]. The performance of plasmas for efficiently holding these processes will depend on plasma parameters, such as the electron density ($n_e$), the electron temperature ($T_e$) and the gas temperature ($T_g$). Therefore, searching of new and easier diagnosis methods to reliably determine these parameter values is becoming a field of increasing interest for the development of plasma applications.

Nowadays one of the most employed techniques for plasma diagnosis is Optical Emission Spectroscopy (OES) which collects atomic and molecular emission from plasmas. Among them, the hydrogen atom is particularly interesting due to its ubiquitous presence in all plasmas. The shape of the line emitted is governed by the values of the physical parameters of the plasma. More specifically, the Balmer alpha line is frequently used [7-9] due to its high intensity. In these works, spectroscopic measurements of this line are employed to obtain the value of the electron density, which is crucial for optimizing applications such as the analysis of volatile species using plasmas as a catalytic medium. This information also aids in the enhancement of new setups like laser-induced optical breakdown plasmas and microwave plasmas sustained at atmospheric pressure. It is worth mentioning that using other lines solely, such as the Balmer beta and gamma lines, do not provide reliable values of this quantity.

While most of the physical models of the different phenomena taking place in the plasma during the emission process provide analytical expressions for the line shapes, either Gaussian or Lorentzian [9-13], this is not the case for the line broadening due to the electric field, generating Stark effect, of the charges interacting with the emitter atom (hydrogen). Specifically, in reference [12], authors discuss how phenomena such as ionic dynamics have a non-negligible effect on the fitting to one Lorentzian profile over the Balmer alpha line.

In the Frequency Fluctuation model (FFM) [14-16] the Stark broadening of the spectral lines is due to a micro-field (local electric field surrounding the emitting atom created by charged particles) with fluctuations (due to collisions with electrons and ions). The first give rise to line splitting while, following Griem [17], three possible regimes can be identified for collisions. The first one, strong and fast collisions, that give rise to a Lorentzian profile of each Stark line with a constructive overlapping between them. The second one, strong and weak collisions, generates a destructive overlapping. And the last one produces a total separation between these Lorentzian profiles. Therefore, the net effect can be described as a superposition of Lorentzian profiles shifted by the micro-field, with constructive or destructive superposition or total separation depending on the nature of the collisions with the charged particles which are governed by the plasma parameters i.e. mass of the ions, ionic and electronic temperature, and density [18].

Within this scheme, line splitting due to the micro-field can be estimated straightforwardly in terms of a model electric field. However, the width of each Lorentzian profile induced by ion and electron collisions is not easy to describe in terms of a simple model. In this sense, the knowledge of Stark broadening experimental data to compare with would be an invaluable resource to tune the model. However clean experimental data on Stark broadening does not exist because the other physical effects, besides Stark broadening, appear in the experimental data as instrumental broadening (which is generally known), van der Waals broadening arising from collisions between the emitting particles and neutral hydrogen atoms—dependent on the plasma's gas temperature—and the Doppler effect, which is related to the thermal motion of the emitters and varies with the emitter's temperature. Separating these effects through deconvolution is a complex computational task that introduces a level of uncertainty that is challenging to quantify. For this reason, theoretical spectral data obtained from realistic Computer Simulations (CS model) by Gigosos et al. [19, 20] including only Stark broadening, have been used in the model proposed in this paper. This CS model is a realistic simulation of hydrogen atoms inside an electrically neutral ensemble of statistically independent charged particles made up of ions and electrons moving within an interaction sphere, thus providing the benchmark results for the spectral lines, which are parametrized in terms of the plasma parameters: electron density $n_e$, electron temperature $T_e$ and gas temperature $T_g$.

In a recent series of works [21-24] these ideas have been tested with the spectral profiles of the $H_\alpha$ line obtained in CS model. For the sake of simplicity, a reduced number of Lorentzian profiles (one, three or five) was employed in the fitting, instead of considering the complete splitting in fifteen lines of this transition due to the Stark effect. Even with this reduced scheme three patterns in the spectra, depending on electron density and temperature, could be identified. Thus zone 1 and zone 2, corresponding to the high and intermediate gas temperature, both ions and electrons induce field fluctuations through impact collisions. The difference between zone 1 and zone 2 is that a constructive/destructive superposition of the splitted profiles is found in the former, while the superposition is practically inexistent in the later. In zone 3, corresponding to the lowest gas temperature region, ions practically not participate in collisions and electrons are the main responsible of collisions.

In this paper, a more realistic model of the Stark broadening of the $H_\alpha$ line in plasmas, considering the full splitting of the transition by including fifteen Lorentzian profiles that correspond to all possible lines, is proposed and implemented. Thus, the number of Lorentzian functions incorporated into the model is only determined by a physical criterion instead of simplicity reasons. This offers an interpretation of the parameters of the different Lorentzian functions included in the model in terms of the features of the local electric field acting over the emitting atom. The model is not as simple as previous ones but a physical picture closer to the processes taking place in the system and a better accuracy on the profiles is achieved. The structure of the paper is as follows: in Section 2, the Stark shifting of $H_\alpha$ line and the CS model is reviewed and the model here proposed is presented; in Section 3, the results of the fitting of data in the ($T_e$, $n_e$, $T_g$) space are analyzed in terms of the different collision contributions; finally, Section 4 is devoted to present the conclusions obtained in this work.

## 2. Theory.

### 2.1. Proposed model.

The CS model data for the Stark profiles are the result of realistic simulations of the emission of the hydrogen atom at different ($T_e$, $n_e$, $T_g$) conditions in plasma. Temperature and ion mass effects are included through the reduced fictitious mass parameter ($\mu_r$), defined as

$$\mu_r = \mu \frac{T_e}{T_g} \tag{1}$$

with $\mu$ the reduced mass between emitter H atoms and the ion (in a.m.u). This parameter accounts for the ion mobility which is governed by both ion mass and gas temperature. For example, for an Ar-H pair, which corresponds to $\mu \sim 1$, a value of $\mu_r = 10$ indicates an electron temperature ten times higher than the gas temperature. In addition, this model does not include fine structure effects, which are negligible for the Hα line at electron densities above 1020 m$^{-3}$ [13].

The data of the simulations are provided at different points of the ($T_e$, $n_e$, $\mu_r$) space.

These results are labelled in terms of a nondimensional parameter, $\rho$, defined, as the ratio between, $R_0$, the mean interparticle distance and, $R_D$, the Debye length, the radius of the sphere where the charged particles exist,

$$\rho = \frac{R_0}{R_D} = \left(\frac{3}{4\pi}\right)^{1/3} \left(\frac{q^2}{\varepsilon_o k_B}\right)^{1/2} \frac{n_e^{1/6}}{T_e^{1/2}} \qquad (2)$$

where $\varepsilon_o$ is the vacuum permeability, $k_B$ the Boltzmann constant and $q$ the electron charge. In Table 1 we show the electronic density and temperature, in terms of the $\rho$ parameter for which the CS model data sets are calculated. For each pair of $n_e$, and $\rho$, seventeen $\mu_r$ values between 0.1 and 10 are considered for the simulations, providing a total of 2992 data files. All these 2992 CS profiles have been used in this work.

| log $n_e$ (m⁻³) | $\rho$ | | | | | | | | | | |
|---|---|---|---|---|---|---|---|---|---|---|---|
| | 0.10 | 0.15 | 0.20 | 0.25 | 0.30 | 0.35 | 0.40 | 0.45 | 0.50 | 0.55 | 0.60 |
| 20.00 | 37508 | 16670 | 9377 | 6001 | 4168 | 3062 | 2344 | 1852 | 1500 | 1240 | 1042 |
| 20.33 | 48444 | 21531 | 12111 | 7751 | 5383 | 3955 | 3028 | 2392 | 1938 | 1601 | 1346 |
| 20.67 | 62568 | 27808 | 15642 | 10011 | 6952 | 5108 | 3910 | 3090 | 2503 | 2068 | 1738 |
| 21.00 | 80809 | 35915 | 20202 | 12929 | 8979 | 6597 | 5051 | 3991 | 3232 | 2671 | 2245 |
| 21.33 | 104369 | 46386 | 26092 | 16699 | 11597 | 8520 | 6523 | 5154 | 4175 | 3450 | 2899 |
| 21.67 | 134798 | 59910 | 33699 | 21568 | 14978 | 11004 | 8425 | 6657 | 5392 | 4456 | 3744 |
| 22.00 | 174098 | 77377 | 43525 | 27856 | 19344 | 14212 | 10881 | 8597 | 6964 | 5755 | 4836 |
| 22.33 | 224856 | 99936 | 56214 | 35977 | 24984 | 18356 | 14054 | 11104 | 8994 | 7433 | 6246 |
| 22.67 | 290413 | 129072 | 72603 | 46466 | 32268 | 23707 | 18151 | 14341 | 11617 | 9600 | 8067 |
| 23.00 | 375083 | 166704 | 93771 | 60013 | 41676 | 30619 | 23443 | 18523 | 15003 | 12399 | 10419 |
| 23.33 | 484438 | 215306 | 121110 | 77510 | 53825 | 39546 | 30277 | 23923 | 19378 | 16014 | 13457 |
| 23.67 | 625676 | 278078 | 156419 | 100108 | 69520 | 51076 | 39106 | 30898 | 25027 | 20684 | 17380 |
| 24.00 | 808092 | 359152 | 202023 | 129295 | 89788 | 65967 | 50506 | 39906 | 32324 | 26714 | 22447 |
| 24.33 | 1043690 | 463862 | 260923 | 166990 | 115966 | 85199 | 65231 | 51540 | 41748 | 34502 | 28991 |
| 24.67 | 1347978 | 599101 | 336995 | 215676 | 149775 | 110039 | 84249 | 66567 | 53919 | 44561 | 37444 |
| 25.00 | 1740981 | 773769 | 435245 | 278557 | 193442 | 142121 | 108811 | 85974 | 69639 | 57553 | 48361 |

**Table 1.** $T_e$ in K for the conditions of $n_e$ and $\rho$ for each $\mu_r$ in CS model [19]

As $\rho < 1$, the electric field around the emitting atom is non-zero giving rise to Stark profile in the transitions. Two different effects take place. First a component of the electric field that is practically constant leads to line shifting around the central wavelength that, for the Balmer alpha line, is given by [24, 25],

$$d_k \ (nm) \approx 10^{-16} \ C \ k \ n_e^{2/3} \qquad (3)$$

where $n_e$ is in m$^{-3}$, $C$, is a shielding coefficient depending on the plasma conditions and the symmetrical $k$-components are $k = 0, \pm1, \pm2, \pm3, \pm4, \pm5, \pm6$ and $\pm8$, where $k = 0, \pm1, \pm5$ and $\pm6$ correspond to $\Delta m = \pm1$ ($\sigma$ polarization) and the others to $\Delta m = 0$ ($\pi$ polarization). The relative intensities of these lines for a given static electric field can be calculated by using atomic wave functions, with values shown in Table 2.

| | σ polarization | | | | π polarization | | | |
|---|---|---|---|---|---|---|---|---|
| k | 0 | ± 1 | ± 5 | ± 6 | ± 2 | ± 3 | ± 4 | ± 8 |
| Relative intensity (%) | 38.81 | 27.37 | 0.23 | 0.26 | 5.15 | 16.29 | 11.88 | 0.01 |

**Table 2.** Relative intensities of the $k$-lines, Eq. (3), shifted by a uniform static electric field.

The second effect is a broadening induced by collisions of charged particles in the plasma, electrons, $\omega_e(n_e, T_e)$ and ions $\omega_i(n_e, T_g)$, with the emitter H atom, provoking a total broadening $\omega(n_e, T_e, T_g)$ of each $k$-component, governed by $n_e$, $T_e$ and $T_g$.

$$\omega(n_e, T_e, T_g) = \omega_e(n_e, T_e) + \omega_i(n_e, T_g) \qquad (4)$$

Hence, the same broadening over each Lorentzian profile can be considered for the widths of the lines. Thus, the model here proposed for the Stark profile of the Balmer alpha line contains a central Lorentzian profile, $L_0$, and seven symmetrical Lorentzian profiles, $L_{\pm 1}$, $L_{\pm 2}$, $L_{\pm 3}$, $L_{\pm 4}$, $L_{\pm 5}$, $L_{\pm 6}$ and $L_{\pm 8}$.

$$P_S(T_e, n_e, T_g; \lambda) = \sum_{\substack{k=-8 \\ |k| \neq 7}}^{8} L_k = \sum_{\substack{k=-8 \\ |k| \neq 7}}^{8} \frac{2a_{|k|}}{\pi} \frac{\omega}{4(\lambda - d_k)^2 + \omega^2} \qquad (5)$$

with, $a_{|k|}$, the area of each profile provides the intensity of the transition, $\omega$, the full width at half intensity maximum (FWHM), the same for all the $k$-components, and because of Equation (3), a linear dependence of the displacements, $d_k$, with $k$ is assumed,

$$d_k = s\,k \qquad (6)$$

with, *s,* the slope, a parameter. Therefore, the model here proposed contains ten temperature and density dependent free parameters, {$s, \omega, a_0, a_k$} (with $k$ = 1, 2, 3, 4, 5, 6 and 8) that are fitted for each ($\rho, n_e, \mu_r$) set of values to the corresponding CS model profile.

Fittings are performed sequentially for each electron temperature and density, and reduced fictitious mass set of data, ensuring that each fitting is entirely independent of the previous one. This method results in a total of 2992 independent curve fittings. Each of the spectral curves is modelled as the sum of fifteen Lorentzian profiles: one central Lorentzian located at $k$ = 0, seven Lorentzian profiles at positive wavelength values, and their seven corresponding symmetric counterparts at negative wavelength values. Consequently, a total of eight Lorentzian profiles are driven during the fitting. The search of the optimum values of the parameters is carried out by using a genetic algorithm [26].

### 2.2. Details of the implemented Genetic Algorithm

To expedite the fitting process, the slopes, *s*, and FWHMs, $\omega$, are restricted to values between 0 and 100, while the areas are between 0 and 1. The solution is represented by a chromosome consisting of a set of genes, where each gene represents a parameter to be adjusted in the theoretical model. In this study, we use a chromosome with 10 genes (slope, FWHM and areas), each encoded in decimal format to facilitate interpretation and manipulation of the parameters. The fitness of each chromosome is evaluated using a Figure of Merit (FoM) that quantifies the quality of the fit between the expected and predicted curves.

$$\text{FoM} = \sum_{i=1}^{n} \frac{(O_i - E_i)^2}{E_i} \qquad (7)$$

where the index *i* runs from 1 to *n*, the number of spectral points defining the CS profiles, $O_i$ is the analytical value obtained from the current value of the parameters at that point and $E_i$ is the value to be fitted. The genetic algorithm was implemented in MATLAB.

For the selection of chromosomes that advance to the next generation, we use the *selectionstochunif* method defined in MATLAB. In brief, the algorithm arranges the parents along a line, where each parent's segment length is proportional to its scaled fitness value. The algorithm then progresses along the line in equal-sized steps. At each step, the algorithm selects a parent from the segment it lands on. The starting point is determined by a uniformly distributed random number that is less than the step size.

Crossover is performed using the *crossoverscattered* operator defined in MATLAB, the function generates a random binary vector. It selects the genes corresponding to 1s in the vector from the first parent and the genes corresponding to 0s in the vector from the second parent. These selected genes are then combined to form the offspring. The crossover probability used is 80%.

We use the *mutationgaussian* operator defined in MATLAB, this operator adds a random number, chosen from a Gaussian distribution to each entry of the parent vector. Typically, the magnitude of the mutation, which is proportional to the standard deviation of the distribution, decreases with each successive generation.

Population size is of 200 chromosomes, the number of generations of 500 and the crossover probability: 80%. The stopping criterion is when the maximum number of generations (1000) or when the FoM stabilizes and does not significantly improve over 50 consecutive generations. The fitting results show satisfactory convergence with a minimum FoM shown in the Figure 1.

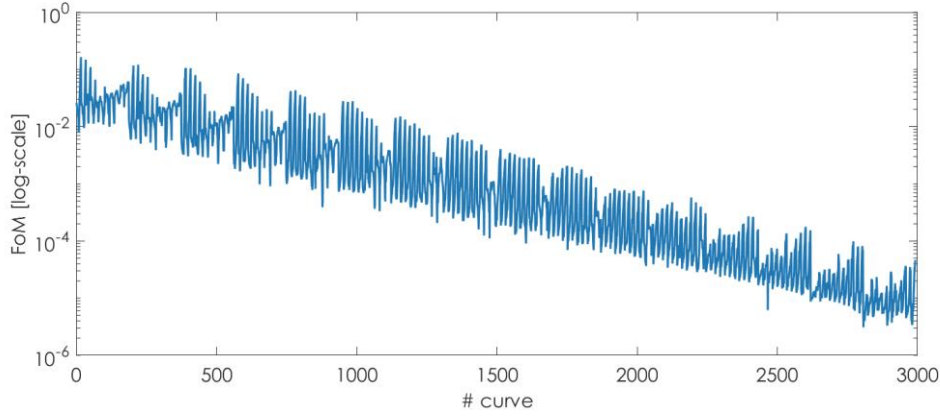

**Figure 1.** FoM value obtained for each of the 2992 evaluated curves.

## 3. Results.

First, in order to illustrate the performance of the present model as compared to previous ones, we show in Figure 2 the profile at a given plasma conditions obtained by using different parameterizations. It is important to note that the main purpose of this work was not to obtain better accuracy by increasing the number of fitting functions, but to build a model based on a physical picture of the effects of local electric field on the emitting atom.

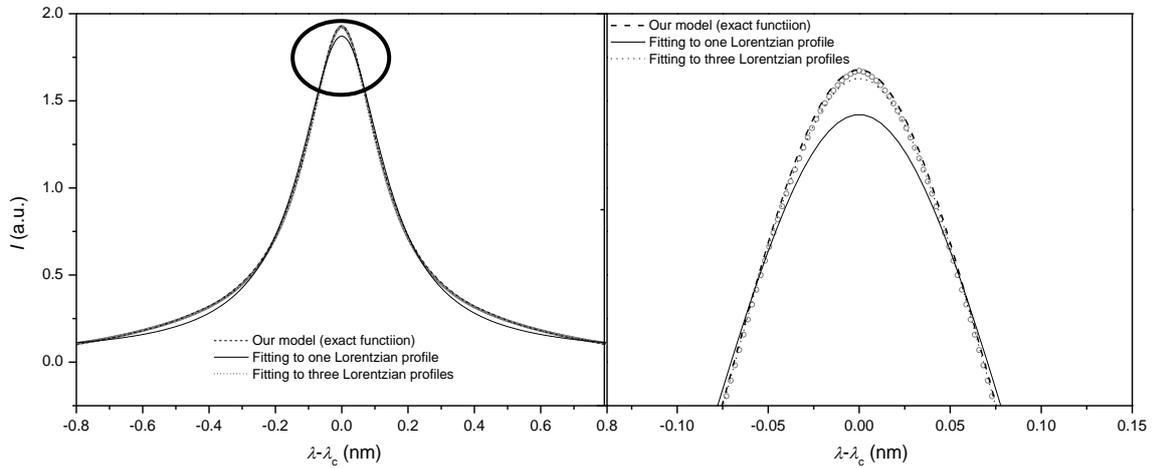

**Figure 2.** Comparison between the fit to one, three or fifteen Lorentzian profiles for $\mu_r = 6$, $T_e = 24984$ K and $n_e = 2.14 \cdot 10^{22}$ m$^{-3}$

Second, we present the results obtained for the fitting parameters of the model profile of this work and analyze the behavior for different plasma conditions. Third, we discuss the different processes taking place in plasma in the light of this analysis. In the supplementary material we provide all the values of the fitting parameters, Eqs. (5) and (6) for the 2992 plasma conditions studied.

In Figure 3 we show the results for the slope, *s*, for some representative values of temperatures and electron densities in the range here studied. The results obtained confirm the hypothesis of the linear dependence of the displacements, Eq. (6). In general, the slope value is very stable except for lower values of the fictitious reduced mass, where it oscillates. The region of oscillations is more relevant as the electron density decreases. Starting from Eq. (3) and using the values of the slope, the shielding parameter of the electric field can be estimated, obtaining values in the range $1.65 \leq C \leq 2.45$.

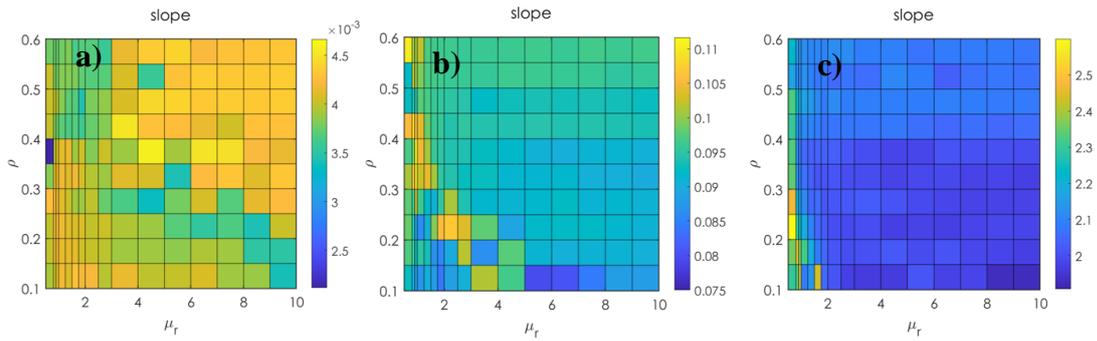

**Figure 3.** Slope parameter for three representative values of the electron density: a) $1 \cdot 10^{20}$ m$^{-3}$; b) $1 \cdot 10^{22}$ m$^{-3}$; c) $1 \cdot 10^{24}$ m$^{-3}$.

In Figure 4 we show the values obtained for the FWHM parameter, $\omega$, for the same electron densities as in Figure 1. The values of $\omega$ decrease with $\mu_r$, except in those regions where the slope fluctuations were observed, where the FWHM also presents fluctuations. Note that in the limit of large $\mu_r$ (small $T_g$) the total width is only due to the electron contribution $\omega\ (n_e, T_e, T_g) \approx \omega\ _e(n_e, T_e)$.

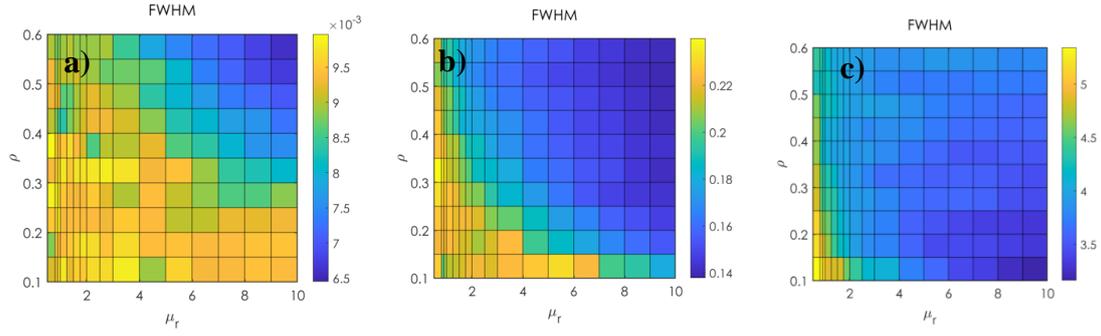

**Figures 4.** FWHM parameter for three representative values of the electron density: a) $1 \cdot 10^{20}$ m$^{-3}$; b) $1 \cdot 10^{22}$ m$^{-3}$; c) $1 \cdot 10^{24}$ m$^{-3}$.

We have found that the onset of stability is $\omega = 2\,s = d_2$. This is illustrated in Figures 5-7, where we plot the slope and FWHM at $\rho = 0.25$ for three different plasma conditions: $n_e = 1 \cdot 10^{20}$ m$^{-3}$, $T_e = 6001$ K; $n_e = 1 \cdot 10^{22}$ m$^{-3}$, $T_e = 27856$ K; and $n_e = 1 \cdot 10^{24}$ m$^{-3}$, $T_e = 129295$ K evaluated for the different fictitious reduced mass here considered. For the first one no stability is found, for the second one stability appears for $\mu_r = 4$ while for the last one, it occurs at $\mu_r = 2$.

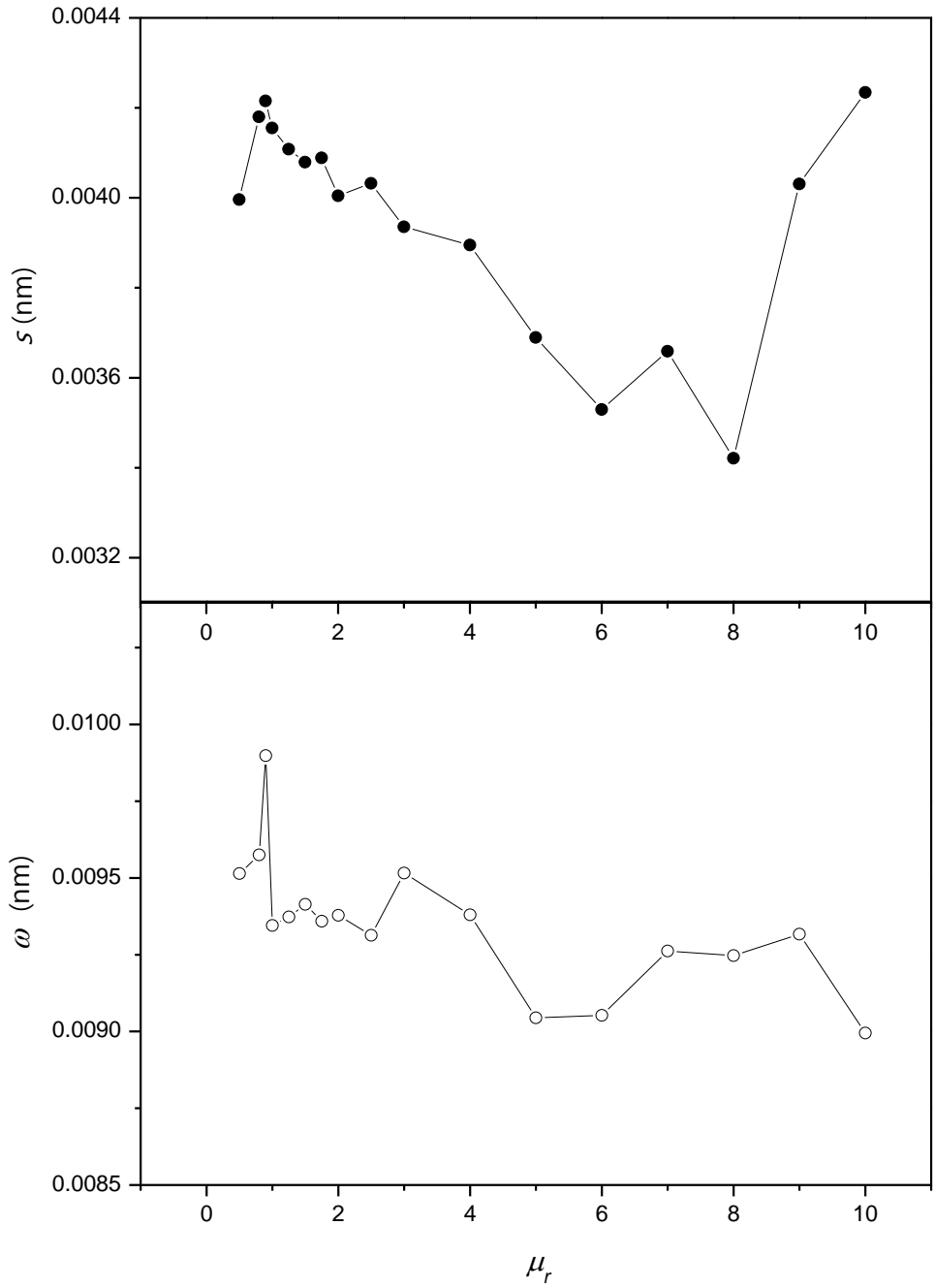

**Figure 5.** Slope (*s*) and width (*ω*) values for the different reduced mass here studied for plasma conditions $\rho = 0.25$ at $n_e = 1 \cdot 10^{20}$ m$^{-3}$, $T_e = 6001$ K for the different $\mu_r$ values considered. The lines are for guiding the eyes.

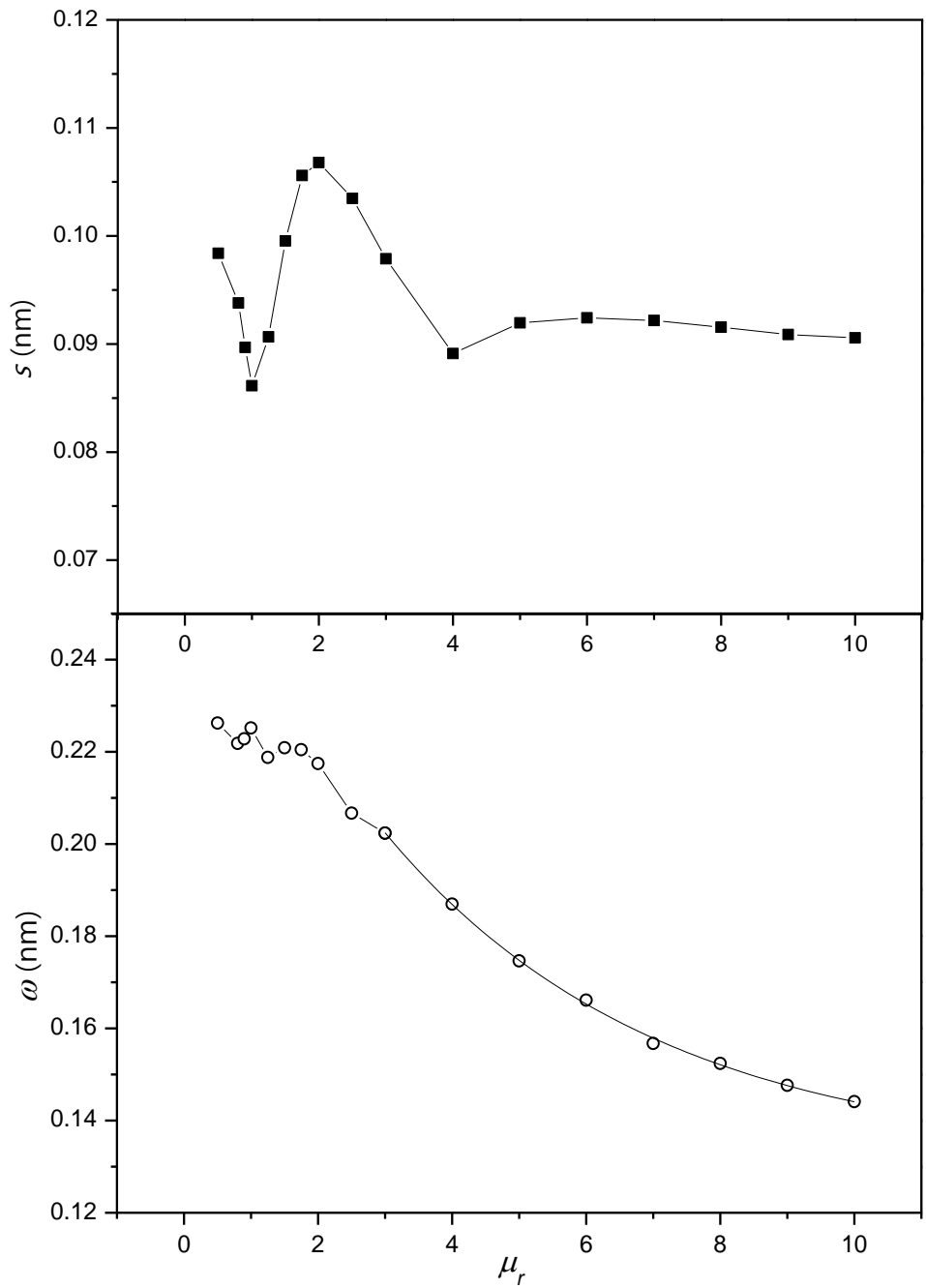

**Figure 6.** Slope (*s*) and width (*ω*) values for the different reduced mass here studied for plasma conditions $\rho = 0.25$ at $n_e = 1 \cdot 10^{22}$ m$^{-3}$, $T_e = 27856$ K for the different $\mu_r$ values considered. The lines are for guiding the eyes.

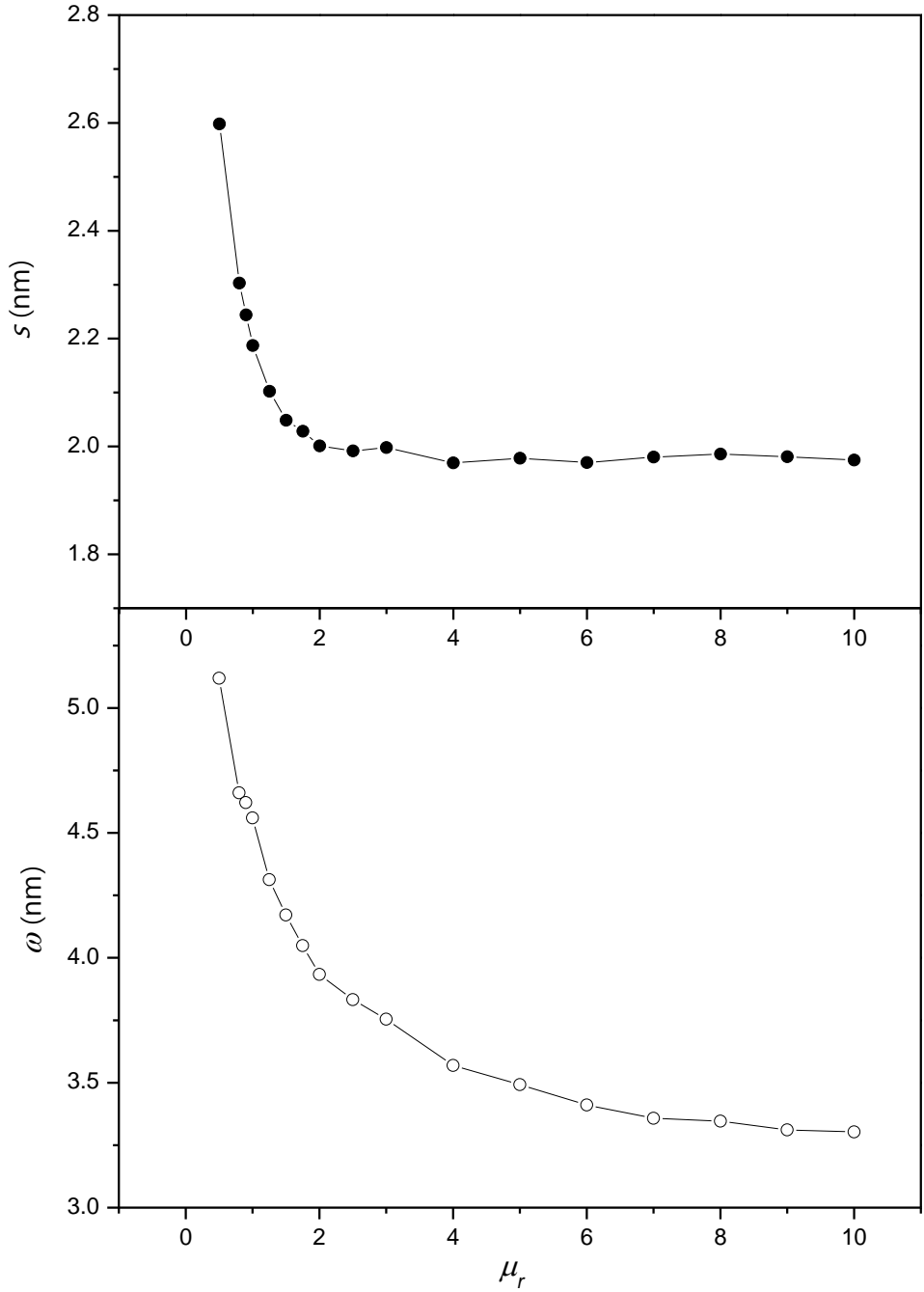

**Figure 7.** Slope (*s*) and width ($\omega$) values for the different reduced mass here studied for plasma conditions $\rho = 0.25$ at $n_e = 1 \cdot 10^{24}$ m$^{-3}$, $T_e = 129295$ K for the different $\mu_r$ values considered. The lines are for guiding the eyes.

Finally, in Figures 8-10 we plot the areas of the first four Lorentzian functions contributing to the profile, $a_0$, $a_1$, $a_2$ and $a_3$ for the same plasma conditions as before.

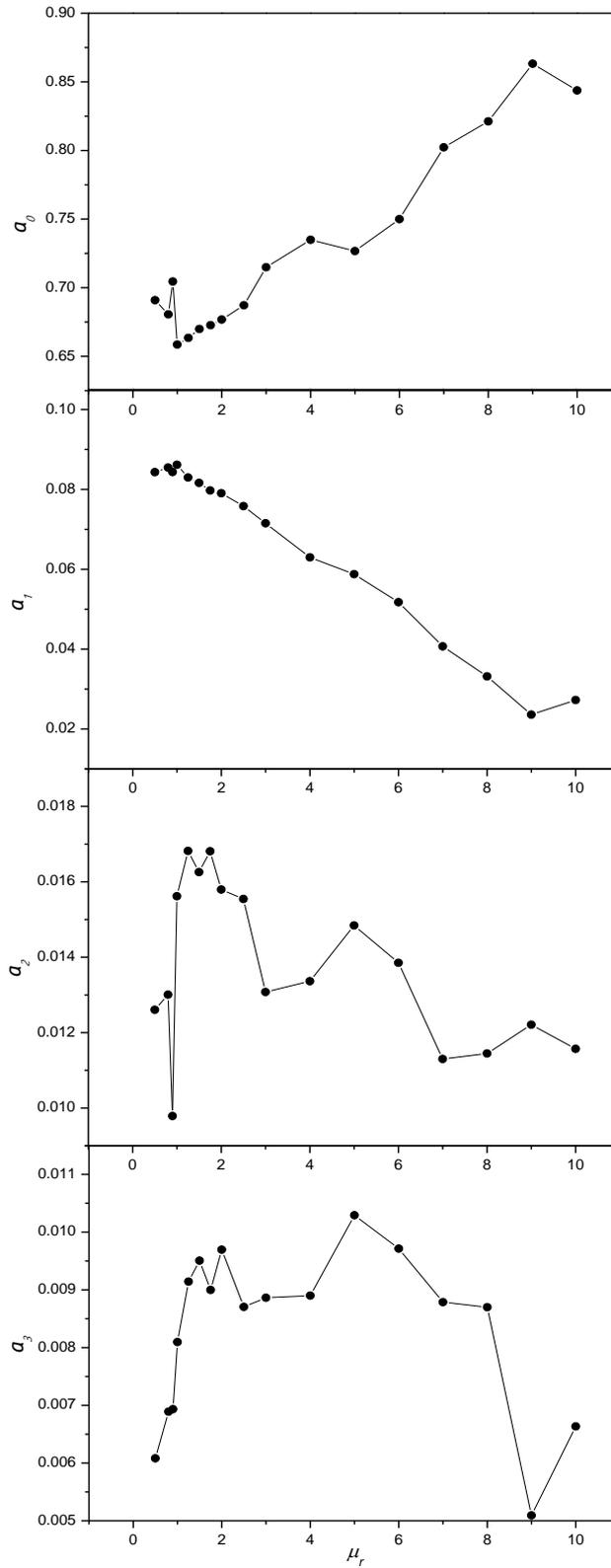

**Figure 8.** $a_0$, $a_1$, $a_2$ and $a_3$ values for the different reduced mass here studied for plasma conditions $\rho = 0.25$ at $n_e = 1 \cdot 10^{20}$ m$^{-3}$, $T_e = 6001$ K for the different $\mu_r$ values considered. The lines are for guiding the eyes.

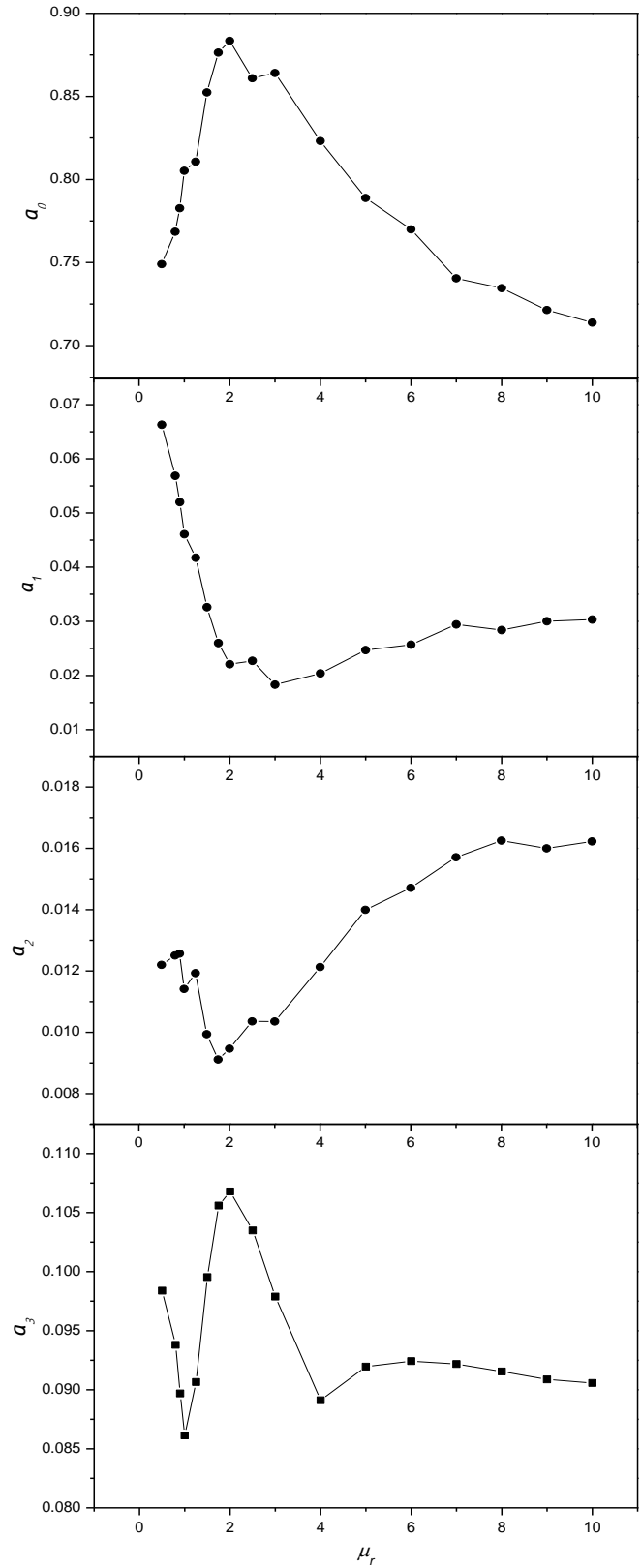

**Figure 9.** $a_0$, $a_1$, $a_2$ and $a_3$ values for the different reduced mass here studied for plasma conditions $\rho = 0.25$ at $n_e = 1 \cdot 10^{22}$ m$^{-3}$, $T_e = 27856$ K for the different $\mu_r$ values considered. The lines are for guiding the eyes.

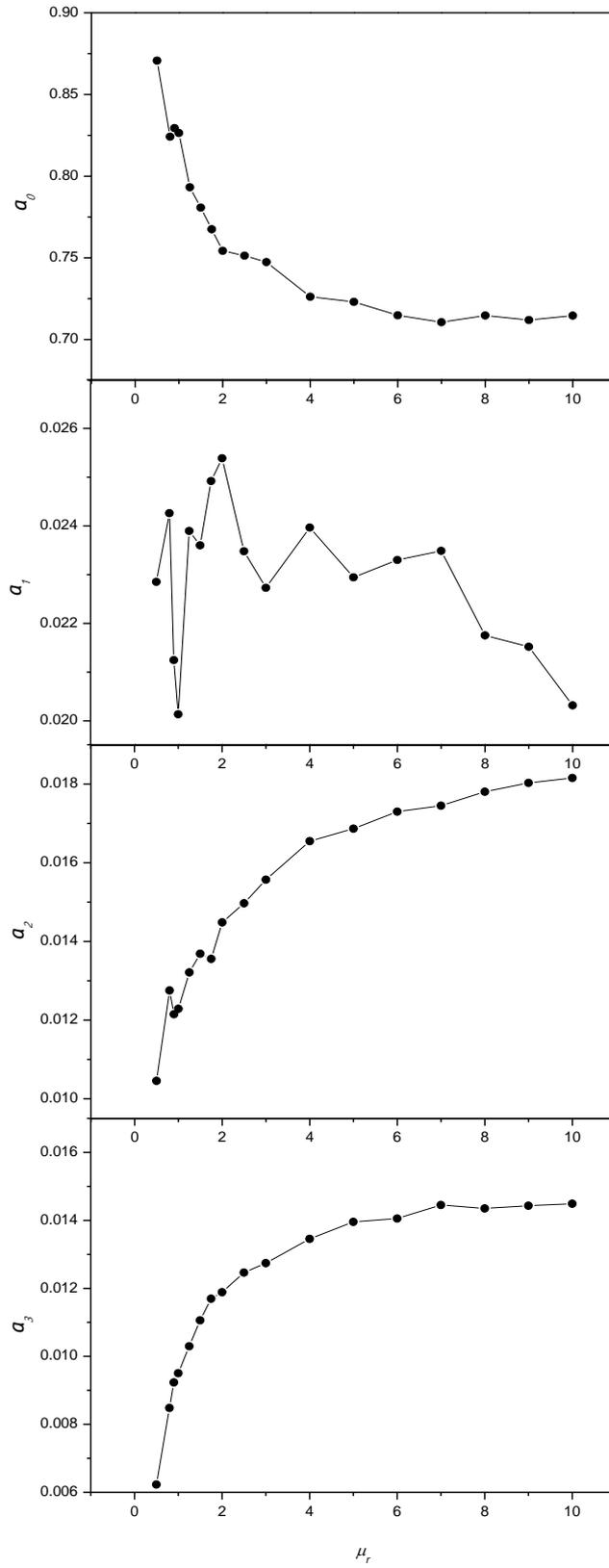

**Figure 10.** $a_0$, $a_1$, $a_2$ and $a_3$ values for the different reduced mass here studied for plasma conditions $\rho = 0.25$ at $n_e = 1 \cdot 10^{24}$ m$^{-3}$, $T_e = 129295$ K for the different $\mu_r$ values considered. The lines are for guiding the eyes.

The areas present a similar behavior as the slopes and the widths, showing a stability region where the areas tend to an asymptotic value. This region of stability corresponds to the larger $n_e$ and $\mu_r$ here considered, where the ion mobility is hindered. The emitting atoms are in an electric field that varies slowly, and the distribution of the quasi-static field is similar for the different plasma conditions. On the other hand, the strong dependence of the parameters with the fictitious reduced mass for low $n_e$ and $\mu_r$ is a consequence of the importance of the movement of the perturber ions on the atom.

At low densities, the areas show a strong dependence with $\mu_r$ at low values of the magnitude. This dependence decreases with increasing fictitious reduced mass and electron density. The points where the behavior of the areas changes are the same as those found for slope and FWHM.

These asymptotic behaviors at higher electron densities and $\mu_r$ imply that ions become more static what translates into greater stability in the slope and a decrease in the total collision contribution (see Eq. 4), but also electron contribution to the areas increases more quickly than ionic.

**Conclusions**

A model for the Stark broadening of the $H_\alpha$ spectrum in plasma in a wide range of plasma conditions is presented. The model is based on the ideas of the Frequency Fluctuation model, providing a framework for understanding Stark broadening in plasmas, attributing it to micro-field fluctuations surrounding emitting atoms due to collisions with charged particles. A full splitting of the transition into fifteen Lorentzian profiles is considered, providing a comprehensive approach to analyzing Stark profiles in plasmas. Computer Simulations are employed as benchmark results for spectral lines, allowing for the parametrization of Stark profiles in terms of ten temperature- and density-dependent coefficients. A fitting method based on genetic algorithms has been employed obtaining very robust values of the fitting parameters for all the plasma conditions considered. The analysis of fitting parameters: the linear coefficient of the displacements, the Full Width at Half Maximum, and areas of Lorentzian profiles, in terms of electron density, temperature, and reduced mass under various plasma conditions, reveals the existence of different regions where there is greater variation in the parameters versus other domains where the parameter values are stable. This behavior is discussed by taking into account the physical scenarios induced by the plasma conditions leading to different ion mobility.

**Acknowledgements**

This work was partially supported by under Grant PID2020-114807GB-I00 by the Spanish MCIN/AEI/10.13039/501100011033.